\documentclass{article}

\usepackage{arxiv}

\usepackage[utf8]{inputenc} 
\usepackage[T1]{fontenc}    
\usepackage{hyperref}       
\usepackage{url}            
\usepackage{booktabs}       
\usepackage{amsfonts}       
\usepackage{nicefrac}       
\usepackage{lipsum}		
\usepackage{graphicx}
\usepackage{epstopdf, epsfig}
\usepackage{amssymb}
\usepackage{amsmath}
\usepackage{natbib}
\usepackage{doi}

\title{Scaling of secondary flows with surface parameters: a linear approach}

\author{ \href{https://orcid.org/0000-0002-5730-4430}{\hspace{1mm}Gerardo Zampino} \\
	Faculty of Engineering and Physical Science\\
	  University of Southampton\\
	Southampton, United Kingdom, SO17 1BJ \\
	\texttt{g.zampino@soton.ac.uk} \\
	\And
	\href{https://orcid.org/0000-0002-6501-6041}{\hspace{1mm}Davide Lasagna} \\
	Faculty of Engineering and Physical Science\\
	  University of Southampton\\
	Southampton, United Kingdom, SO17 1BJ \\
    \And
	 \href{https://orcid.org/0000-0001-9817-0486}{\hspace{1mm} Bharathram Ganapathisubramani} \\
	 Faculty of Engineering and Physical Science\\
	   University of Southampton\\
	 Southampton, United Kingdom, SO17 1BJ 
}

\renewcommand{\shorttitle}{\textit{arXiv} Template}

\begin{document}
\maketitle

\begin{abstract}
	Secondary flows are generated when a lateral variation of the topography, such as streamwise aligned ridges, is imposed to a turbulent wall-bounded flow. In this case, the flow field is characterized by vortical structures developing along the streamwise direction known as Prandt's vortices of the second kind \citep{prandtl1952}. As demonstrated in previous experimental and numerical works, the strength and flow organization of these turbulent structures depend on the ridge shape. In this paper, the effect of the ridge geometry on the generation of secondary currents is investigated using the linearised RANS-based model proposed by \citet{zampino2022}. Here, symmetric channels with rectangular, triangular and elliptical ridges are studied and the secondary flows are compared in order to highlight the main differences and similarities. The analogies of the flow organisation between the three geometries suggest that the secondary currents do not depend on the ridge shape when the ridges are small and isolated, and the strength of secondary flows collapses when properly scaled with the mean ridge height. Finally, the generation of secondary flows and the effect of the ridge shape on the flow organisation is studied in detail for the complex geometries. In particular, for trapezoidal ridges, we observed that tertiary flows emerge for the ridges where the scaling behaviour does not hold.
\end{abstract}


\section{Introduction}
\label{sec:introduction}
The generation of secondary currents over a surface with heterogeneous attributes, such as a lateral variation of the roughness height or of the wall topography, is largely studied in literature. These vortices are known as Prandtl's vortices of the second king \citep{prandtl1952}. Since the first experiments conducted by \citet{hinze1967}, these turbulent structures have gained increasing importance in industrial applications because secondary flows alter and modify the performances of fluid dynamics surfaces such as the transport properties of the wall-bounded flows \citep{volino2011,mejia2013,vanderwel2015,Hwanglee2018,medjnoun_vanderwel_ganapathisubramani_2020,zampiron2020}, the heat transfer \citep{stroh2020} and the aerodynamic performances \citep{jimenez2004,mejia2013}. For the surfaces with topographical heterogeneity, with alternating regions of high/low relative elevation \citep{Hwanglee2018,medjnoun_vanderwel_ganapathisubramani_2018,medjnoun_vanderwel_ganapathisubramani_2020,castro2020}, the flow organisation is characterised by alternating high-momentum pathways (HMPs) and low-momentum pathways (LMPs) corresponding to a downwash/upwash motion respectively \citep{barros2014,willigham2014}. These structures were observed both experimentally \citep{anderson2015} and numerically \citep{stroh2016,chung2018}. It has been shown that the lateral intensity of HMPs and LMPs and of the associated vortical structures can depend on the shape of the ridges \citep{medjnoun_vanderwel_ganapathisubramani_2020} but how the geometry affects the generation mechanism of secondary structures and their strength is not fully clear. 
Some authors \citep{wu2007,castro2020} observed that, in some cases, secondary flows over different geometries appear to be similar and this suggests a possible scaling of the vortical structures with a geometrical property of the surface. In particular, \citet{castro2020} concluded that the strength of the secondary flows developing over rectangular ridges depends on the ratio between the ridge spacing and width and the flow organization is independent of the spacing when scaled with the channel height. 
Other scaling  geometrical parameters have been proposed in literature, such as the ratio between the wetted area above and below the ridge mean height by \citet{medjnoun_vanderwel_ganapathisubramani_2020}.
In this paper, we address these questions by using a tool based on linearised Reynolds-Averaged Navier-Stokes (RANS) equations as proposed by \citet{zampino2022}. The authors developed a tool for the rapid prediction of the secondary flows using the RANS equations coupled with the Spalart-Allmaras (SA) turbulence model \citep{spalart1994} for the eddy viscosity profile necessary to close the linear system. For the small perturbation assumption, the wall modulation is modelled using linearised boundary conditions, following the methodology proposed by \citet{russoluchini2016}. The tool was recently applied to study the generation of secondary flows in symmetric channels with harmonic wall modulation and rectangular ridges \citep{zampino2022}. Although the linearised model predictions agrees with other available data in literature, the authors observed that it cannot replace DNS or experiments. The approach proposed by \citet{zampino2022} assumes a steady streamwise-independent perturbations and hence it cannot capture highly unsteady behaviour \citep{vanderwel2019} and the meandering phenomena \citep{hutchins2007,kevin2019,zampiron2020}. Secondly, surfaces with prominent ridges cannot be satisfactorily modelled using linearised boundary conditions and the upwards deflection operated by the flanck of the ridges is not predicted. 

The overall purpose of this paper is using this tool to contrast and compare the most common studied ridge geometries as elliptical, triangular and rectangular ridges. 
In addition, the evolution of the flow organisation and the generation of tertiary flows are also discussed for complex geometries in order to provide a better description of the generation mechanisms of these structures as a function of the properties of the ridge shape. The trapezoidal ridges have been studied because this shape combines the properties of the rectangular and triangular ridges.    
A summary of the linearised model proposed in our previous work is reported in section \ref{sec:RANS}. The approach is firstly applied for the description of the flow organisation over rectangular, elliptical and triangular ridges in section \ref{sec:prediction} while the discussion about the scaling of the strength of secondary flows is reported in section \ref{sec:scaling}. The analysis of the flow organisation over trapezoidal ridges and the discussion about the generation mechanism of the tertiary flows are reported in section \ref{sec:trapezoidal}. Finally, conclusions are reported in section \ref{sec:conclusions}

\section{Linearised RANS model}
\label{sec:RANS}
A pressure-driven channel with streamwise aligned ridges is studied. A symmetric channel configuration is here considered. In the following, the streamwise, wall-normal and spanwise directions,  normalized with the channel half height $h$, are identified by the cartesian coordinates $(x_1,x_2,x_3)$ respectively. The coordinate system is centred on the channel mid-plane. 
The flow is governed by the Reynolds-averaged continuity and the momentum equations for the velocity components $(u_1,u_2,u_3)$ scaled with the friction velocity $u_\tau=\sqrt{\tau_w/\rho}$, with $\tau_w = h \Pi$ the mean wall friction and where $\Pi$ is the constant pressure gradient. Given these definitions, $Re_\tau=u_\tau h/\nu$ is the friction Reynolds number. Reynolds-averaging produces the mean velocity $\overline{u}_i$ and the fluctuation $u'_i$.
The nondimensional Reynolds-averaged continuity and momentum equations are
\begin{align}
    \frac{\partial \overline{u}_i}{\partial x_i} =&\;0,\\
    \overline{u}_j\frac{\partial \overline{u}_i}{\partial x_j}=&- { \delta_{i1}}+\frac{1}{Re_\tau}\frac{\partial^2 \overline{u}_i}{\partial x_j^2}-\frac{\partial \overline{u_i'u_j'}}{\partial x_j}.
   \label{eq:RANS}
\end{align}
We assume that the mean flow is streamwise-independent (i.e. $\partial(\cdot)/\partial x_1 \equiv 0$) when considering flow structures that develop over streamwise-independent ridges. Hence, the mean pressure can be eliminated by employing
a streamwise velocity/streamfunction formulation, where the streamfunction $\overline{\psi}$ satisfies $\nabla^2 \overline{\psi}=\overline{\omega}_1$ with 
\begin{equation}
    \overline{\omega}_1= {\frac{\partial \overline{u}_3}{\partial x_2}-\frac{\partial \overline{u}_2}{\partial x_3}}
    \label{omegadefinition}
\end{equation} 
being the streamwise vorticity. 

A sketch of the ridge geometries considered in this paper is reported in figure \ref{fig:topology}(a,b,c), where $S$ is the spacing between the ridges and $W$ is the ridge width. The duty cycle $DC$ is defined as $W/S$. We introduce a unitary periodic modulation of the surface in the spanwise direction as given by the function $f(x_3)$. The bottom physical surface is thus placed at $x_2=-1+\epsilon f(x_3)$ while the upper surface is symmetric. The proposed methodology is based on decomposing the flow field into a homogeneous flow over the flat channel and the flow perturbation induced by the ridges. Defining $\epsilon$ as the peak-to-peak distance of the ridge, we expand any time-averaged flow quantity $q$ using a Taylor series in $\epsilon$ as   
\begin{equation}
    q(x_2,x_3)=q^{(0)}(x_2)+\epsilon q^{(1)}(x_2,x_3)+\mathcal{O}(\epsilon^2),
    \label{taylorexpansion}
\end{equation}
where the zero-order term $q^{(0)}$ is the base flow solution in the flat channel and the first-order term $q^{(1)}$ is the flow response per unit of ridge height. Following \citet{russoluchini2016}, in the limit of small ridges, we assume that the intensity of the secondary currents is proportional to $\epsilon$. Thus, in the definition (\ref{taylorexpansion}) higher order terms in $\epsilon$ are neglected.   
Substituting the Taylor expansion (\ref{taylorexpansion}) in the governing equations written using the streamwise velocity $u_1$ and streamfunction $\psi$ formulation. The overbar that is commonly used in literature for the quantities scaled in inner units, it here omitted for clarity. Therefore, considering the terms at order one in $\epsilon$, we obtain
\begin{align}
\displaystyle
-\frac{\partial \psi^{(1)}}{\partial x_3} \Gamma &\!=\!\frac{1}{Re_\tau}\!\left( \frac{\partial^2 }{\partial x_2^2}\!+\!\frac{\partial^2 }{\partial x_3^2}\right) u_1^{(1)}\!+\!\frac{\partial \tau_{12}^{(1)} }{\partial x_2}\!+\!\frac{\partial \tau_{13}^{(1)} }{\partial x_3},
    \label{OSfinal}\\
    \displaystyle
    \begin{split}
        0&\!=\!\frac{1}{Re_{\tau}}\!\left( \frac{\partial^2 }{\partial x_2^2}\!+\!\frac{\partial^2}{\partial x_3^2}\right)^2\! \psi^{(1)}\!\\&+\!\frac{\partial^2}{\partial x_2 \partial x_3} \left( \tau_{33}^{(1)}\!-\!\tau_{22}^{(1)}\right)\!+\!\left( \frac{\partial^2}{\partial x_2^2}\!-\!\frac{\partial^2}{\partial x_3^2}\right)\! \tau_{23}^{(1)},
    \end{split}
\label{psiequation}
\end{align}
where $\Gamma$ is the zero-order streamwise velocity wall-normal gradient and $\tau_{ij}^{(1)}$ is the Reynolds stress tensor perturbation. Similarly, to close the system of equations (\ref{OSfinal}) and (\ref{psiequation}), the SA transport equation for the perturbation of the eddy viscosity $\nu_t^{(1)}$ induced by the ridges is linearized (here omitted for brevity). 

The tensor $\tau_{ij}^{(1)}$ must be expressed as a function of $\nu_t^{(1)}$ and the other mean quantities. As already discussed in literature of noncircular ducts \citep{perkins1970,bottaro2006}, when the linear Boussinesq's hypothesis is used, no secondary flows can be predicted because the streamwise momentum equation (\ref{OSfinal}) and the streamfunction equation (\ref{psiequation}) are decoupled. Hence, a nonlinear Reynolds stress model is necessary for the correct prediction of anisotropic stresses that are the source of the secondary flows. Many approaches have been described in the literature \citep{speziale1991,speziale1991b, liencubic}.
In this work, we used the Quadratic Constitutive Relation (QCR) nonlinear model that was introduced by \citet{spalart2000}. The QCR model contains simple terms proportional to the product of the rotation and the strain tensors. This model was recently utilised by \citet{spalart2018} to predict the high-Reynolds number asymptotic properties of secondary flows in square and elliptical ducts, providing a good approximation of the secondary vortical flow topology and of the wall friction coefficient. In the QCR, the Reynolds stresses become   

\begin{equation}
    \tau_{ij}=\tau_{ij}^{L}-C_{r1}\left[ O_{ik}\tau_{jk}^{L}+O_{jk}\tau_{ik}^{L}\right], \label{taunonlinear}
\end{equation}
where  $O_{ik}$ is the normalized rotation tensor
\begin{equation}
O_{ij}  = \frac{2W_{ij}}{\sqrt{\displaystyle  {\frac{\partial \overline{u}_m}{\partial x_n} \frac{\partial \overline{u}_m}{\partial x_n}}}}, \quad \mathrm{with}\quad
W_{ij} = \frac{1}{2}\left( \frac{\partial \overline{u}_i}{\partial x_j}-\frac{\partial \overline{u}_j}{\partial x_i}\right).
\label{oijdefinition}
\end{equation}
and $\tau_{ij}^{L}=\nu_t S_{ij}$ is the linear Reynolds stress tensor from the Boussinesq's approximation with $S_{ij}$ the mean velocity gradient tensor and $\nu_t$ the eddy viscosity.  The constant $C_{r1}=0.3$ is calibrated to match the anisotropy in the outer region over wall-bounded turbulent flows following \citet{spalart2000}. 

The one-equation SA turbulence model \citep{spalart1994} is used as closure model for the eddy viscosity $\nu_t$. The eddy viscosity $\nu_t$ is related to the modified eddy viscosity $\tilde{\nu}$ by the formula 
\begin{equation}
\nu_t=\tilde{\nu} f_{v1}    
\end{equation}
where 
\begin{equation}\label{eq:fv1_definition}
    f_{v1}=\frac{\chi^3}{\chi^3+c_{v1}^3}
\end{equation} 
and $\chi=Re_{\tau} \tilde{\nu}$. The transport equation for the perturbation of the modified eddy viscosity $\tilde{\nu}^{(1)}$ is well-known as 
\begin{eqnarray}
     -\frac{\partial \psi^{(1)}}{\partial x_3} \frac{\partial \tilde{\nu}^{(0)}}{\partial x_2}& =& \frac{1}{\sigma}\left( \frac{1}{R_{e_\tau}}+\tilde{\nu}^{(0)}\right) \left( \frac{\partial^2 }{\partial x_2^2}+\frac{\partial^2}{\partial x_3^2}\right) \tilde{\nu}^{(1)}+\frac{1}{\sigma}\frac{\partial^2 \tilde{\nu}^{(0)}}{\partial x_2^2}\tilde{\nu}^{(1)}  \nonumber\\  &+&\frac{1}{\sigma}(2+2 c_{b2}) \frac{\partial \tilde{\nu}^{(0)}}{\partial x_2} \frac{\partial \tilde{\nu}^{(1)}}{\partial x_2} +c_{b1}\tilde{\nu}^{(0)} \tilde{\mathcal{S}}^{(1)}+c_{b1} \tilde{\nu}^{(1)} \tilde{\mathcal{S}}^{(0)}  \nonumber \\ &-2& \tilde{\nu}^{(0)} c_{w1} f_w^{(0)} \frac{ \tilde{\nu}^{(1)} d^{(0)}- \tilde{\nu}^{(0)}d^{(1)}}{d^{(0)}\,^3}-
    c_{w1} f_w^{(1)} \left( \frac{\tilde{\nu}^{(0)}}{d^{(0)}}\right)^2.
    \label{eq:SAfirstorder1}   
\end{eqnarray}
where $\tilde{S}$ is defined as 
\begin{equation}
 \tilde{\mathcal{S}}=\sqrt{2W_{ij} W_{ij}}+ \displaystyle \frac{\tilde{\nu}}{k^2 d^2} f_{v2} \quad \mathrm{with}  \quad f_{v2}= 1- \displaystyle \frac{\chi}{1+ \chi f_{v1}}.
\end{equation}
A detailed description of the other terms in the equation \ref{eq:SAfirstorder1} is reported in \citet{zampino2022}.

It is worth noting that previous studies that that have utilised linearised RANS equations to describe the transient growth of vortical structures in turbulent channels \citep{delalamo2006, pujals2009} have used analytical eddy-viscosity profiles \citep{cess1958, reynolds1972}. In the present paper, the eddy viscosity distribution in the modulated geometry and capture transport effects, we use an eddy viscosity transport model \citep{spalart1994}, initially developed for attached shear flows. 

The effect of the ridges is introduced using linearised boundary conditions \citep{luchini2013,busse2012}.  Expanding the velocity near the surface in a Taylor series in $x_2$ and enforcing the no-slip condition at the physical surface, the streamwise velocity component at the lower domain boundary is given by the inhomogeneous boundary condition 
\begin{equation}
      \displaystyle 
      \left. u_{1}^{(1)} \displaystyle \right|_{x_2=-1} +  f(x_3) \left.\frac{\partial u_1^{(0)}}{\partial x_2}\right|_{x_2=- 1}  = 0, \label{u1bcs}
\end{equation}
i.e.~the perturbation velocity at the boundary of the numerical domain is proportional to the wall-normal gradient of the streamwise velocity in the plane channel. When substituting the definition (\ref{taylorexpansion}) in the equation (\ref{u1bcs}) and considering only the terms at order one, the streamwise velocity perturbation becomes
\begin{equation}
     u_{1}^{(1)}(x_2 = -1) = -  f(x_3) \left.\frac{\partial u^{(0)}}{\partial x_2}\right|_{x_2=- 1} = -  f(x_3) Re_{\tau} , \label{eqn:bcsfinal}
\end{equation}
while $u_{3}^{(1)}(x_2=-1) = 0$ and $u_{2}^{(1)}(x_2=-1) = 0$. Similarly, inhomogeneous boundary conditions are derived for the perturbation eddy viscosity $\nu_t^{(1)}$ which vanishes at the physical surface.
Finally, any periodic wall modulation with an unitary ridge height is modelled by the zero-mean modulation 
\begin{equation}
    f(x_3)=\sum_{n=1}^{\infty} f^n \cos(n k_3 x_3),
    \label{eq:fourier}
\end{equation}
where $f^n$ is the amplitude of the n-th wavenumber mode and $k_3=2 \pi/S$ is the fundamental wavenumber. 
The flow field over any complex geometry can be reconstructed following the superposition principle. Since the first-order equations are linear, a complex solution is given by the constructive (or destructive) interference of solutions at independent wavenumbers. The entire methodology is extensively described in \citet{zampino2022}. 

\section{Linearised prediction of the flow organisation}
\label{sec:prediction}
\begin{figure}
	\centering
	\includegraphics[width=\textwidth]{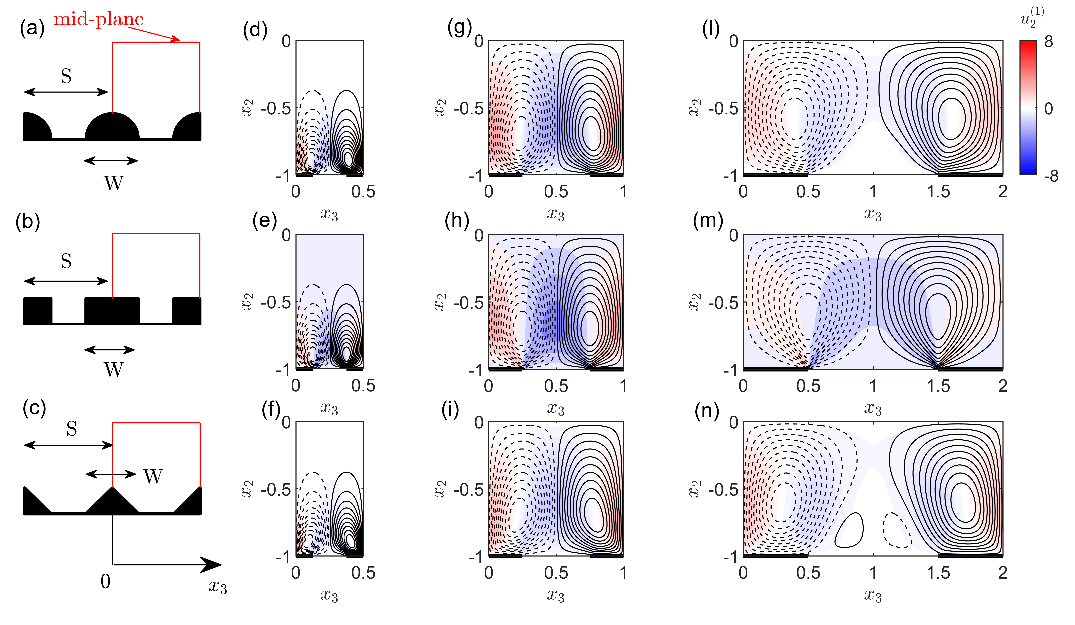}
	\caption{Contour lines of the perturbation streamfunction $\psi^{(1)}$ at $Re_\tau=1000$ and for elliptical ridges (top panels), rectangular ridges (central panels) and triangular ridges (bottom panels). Dashed lines are used for negative values. A sketch of the cases studied is reported in panels (a,b,c). The flow organization is shown for the bottom half channels in the region delimited by the red lines. The colour map of the wall-normal velocity perturbation $u_2^{(1)}$ is also reported to better display the flow organization above the ridges. The duty cycle $DC=0.5$ for all cases while the width varies from 0.25 (left column) 0.5 (central column)  and 1 (right column).   To help the reader, a simplified representation of the ridges is reported at the bottom of each plot (bold black lines). }
	\label{fig:topology}
	\end{figure}
Firstly, secondary flows are predicted over elliptical, rectangular and triangular ridges at $Re_\tau=1000$ with duty cycle $DC=W/S=0.5$ and $W=0.25$ (left panels), $0.5$ (central panels) and $1.0$ (right panels) in figure \ref{fig:topology}. Contours of the perturbation streamfunction $\psi^{(1)}$ and the colour map of the wall-normal velocity perturbation $u_2^{(1)}$ are also reported to better display the flow organization. Because  symmetric channel configurations are studied, subjected to a periodic wall modulation, only the bottom half-channel and a single ridge period is here shown. 

A single Reynolds number is considered. From our previous work \citep{zampino2022}, we observed that the Reynolds number affects only the strength of the secondary flows but the flow organization is mostly unaffected. More specifically, for high Reynolds number, the solution of the present model becomes Reynolds invariant because of the turbulence model used. In fact, the Spalart-Allmaras model \citep{spalart1994} is built in order to obtain a collapse of the eddy viscosity profile in the logarithmic layer for high Reynolds numbers. As a consequence, the eddy viscosity profile and the Reynolds stresses are asymptotically Reynolds number independent when scaled with the friction velocity. For this reason, the results for $Re_\tau=1000$ were here reported and they are representative of the secondary flows generated by the ridges also at high Reynolds numbers.

For small spacing (left column), vortical structures predicted for all three geometries show an upwelling (downwelling) motion above the ridges (inside the troughs). The secondary vortices occupy only about a quarter of the channel height and they are similar in size and strength for all three geometries considered. Although some differences are predicted in the very near-wall region where the ridge geometry affects the local flow organization, these differences are weak and negligible. For increasing ridge width W, the vortices grow in size and strength until they occupy the entire channel half-height (central column). Some differences are here evident in the strength of secondary flows and stronger downwash velocity is predicted at the gap centre for rectangular ridges. 
For $W=1$, very large turbulent structures are observed for all geometries (right column). In particular, for rectangular ridges, the secondary flows are locked at the ridge edge due to the strong discontinuity introduced by the ridge geometry. The same feature is also observed for elliptical ridges where the secondary flows develop in proximity of the ridge edge. By contrast, for triangular ridges, the secondary currents occur at the flank of the ridge. For this case, tertiary flows at the centre of the trough are predicted only for triangular ridges, and produce a local change of the flow direction. For specific spacing and widths, tertiary flows can also be observed at the centre of the trough for the other geometries when the gap between the ridges is large enough to allow the turbulent structures to fully develop. In fact, for large spacing, the vortical structures reach their maximum size and a further increase of the spacing allows tertiary flows to emerge. For similar reasons, tertiary flows can be predicted over the rectangular and elliptical ridges when the ridge width is large enough. By contrast, no tertiary flows over the ridge are observed for the triangular shape because the deflection of the spanwise velocity component induced by the flank of the ridges is weaker.

\section{Scaling of the strength of secondary flows} \label{sec:scaling}
\begin{figure}
    \center
	\includegraphics[width=1\textwidth]{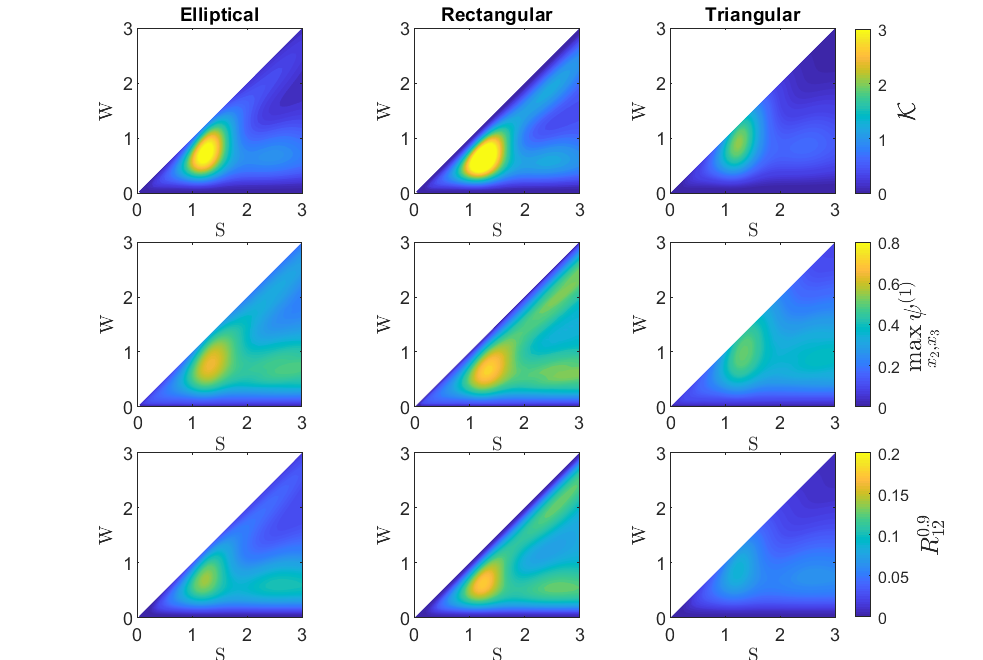}
	\caption{Maps of the kinetic energy density $\mathcal{K}$ (top panels), the maximum of the streamfunction $\max_{x_2,x_3} \psi^{(1)}$ (central panels) and dispersive stress $\hat{R}_{12}^{0.9}$ (bottom panels) are reported as a function of the ridge spacing $S$ and ridge width $W$ for $Re_{\tau}=1000$ for elliptical ridges (left column), rectangular ridges (central column) and triangular ridges (right column).}
	\label{figure1}
	\end{figure}

The strength of secondary flows is not uniquely defined and different authors use different quantities to define the strength of these turbulent structures. As proposed in our previous work \citep{zampino2022}, the secondary flows over rectangular ridges can be characterised using the kinetic energy density 
$$\mathcal{K}=\frac{1}{4S}\! \int_{-1}^{1} \int_{0}^{S} \left[ u_2^{(1)}(x_2, x_3)^2 + u_3^{(1)}(x_2, x_3)^2\right ] \,\mathrm{d}x_3 \,\mathrm{d}x_2,$$ 
and the maximum of the streamfunction $\max_{x_2,x_3} {\psi^{(1)}(x_2,x_3)}$. In addition, many experimental and numerical works,use dispersive stresses $$\sigma_{ij}(x_2,x_3)= u_i^{(1)}(x_2,x_3)u_j^{(1)}(x_2,x_3)$$ to study the generation of secondary flows. In order to characterize their global strength using a single scalar quantity, we also introduce the integral 
\begin{equation}
    R_{ij}^l=\int_{0}^{S}\int_{-l}^{l} \sigma_{ij}(x_2,x_3) \text{d} x_2 \text{d}x_3,
    \label{eq:Rij}
\end{equation}
where we take $l=0.9$ to discard the contribution in the near-wall region. 
Maps of the kinetic energy density $\mathcal{K}$ (top panels), of the maximum of the streamfunction $\max{\psi^{(1)}}$ (central panels) and the quantity $R_{12}^{0.9}$ (bottom panels) are plotted in figure \ref{figure1} for $Re_\tau=1000$, as a function of $S$ and $W$ for these geometries. 
In a first approximation, the three quantities we introduced for the analysis of the strength of the secondary flows are equivalent and the predictions about the amplification of the turbulent structures are unaffected by the quantity chosen. Thus, the maps of figure \ref{figure1} only depends on the ridge shape studied and on the Reynolds number.  

Hereinafter, more general considerations are reported for all cases studied. The linear model predicts i) a region of high amplification at $(S,W) \approx (1.25, 0.67)$ where the strength of the secondary flows is maximum for all quantities considered (small difference in the position of the amplification geometry are observed as a function of the ridge shape), and ii) high amplification along a line at constant $W\approx0.67$ for increasing spacing. However, the peak values across the three cases are different (stronger for rectangular ridges), confirming that the strength of secondary currents depends on the ridge geometry. For high spacing $S$ or width $W$, all quantities considered 
for rectangular ridges also displays a second amplification peak not visible for the other geometries and corresponding to the maximum strength of the tertiary flows developing at the ridge centre (or trough centre). This suggests that tertiary flows are generated in different configurations in terms of $S$ and $W$, depending on the ridge shape. A complete analysis of the generation of tertiary flows as a function of the ridge shape and the effect of the ridge shape is reported in the section \ref{sec:trapezoidal} where the flow organisation is plotted for more complex shapes. 

Despite the difference in the strength of the response, the flow organization observed for the narrow ridge case $W=0.2$ shown in figure \ref{fig:topology} is similar for all geometries, suggesting that the cross-stream velocities and dispersive stresses might be scaled using an appropriate geometrical parameter. In this particular regime, one can assume that the surface perturbation generating the secondary currents is localised in a narrow region at the wall and the effect of the ridge is proportional to its cross-sectional area but not its geometry. For this reason, we introduce the mean ridge height $\overline{H}$ as the ratio between the cross-sectional area and the spacing $S$ and use this quantity to scale the perturbation velocity field.

\begin{figure}
    \center
	\includegraphics[width=1\textwidth]{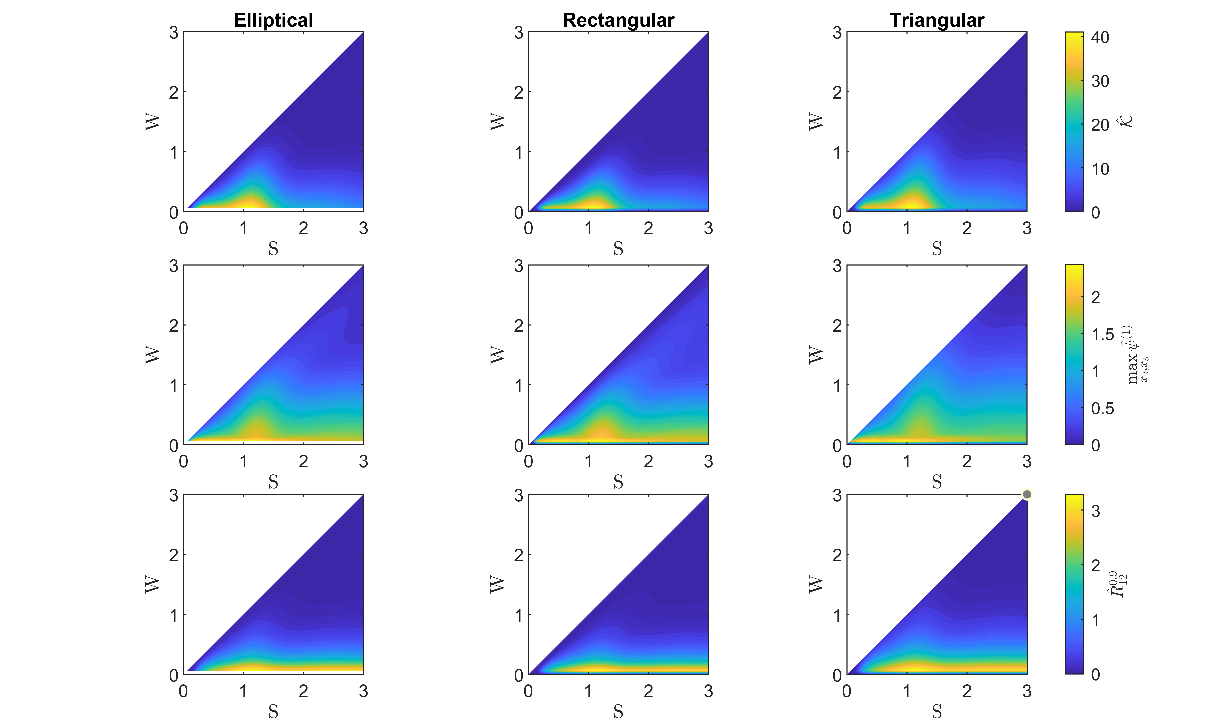}
	\caption{Maps of the scaled kinetic energy density $\mathcal{\hat{K}}$ (top panels), the scaled maximum of the streamfunction $\max_{x_2,x_3} \hat{\psi}^{(1)}$ (central panels) and dispersive stress scaled $\hat{R}_{12}^{0.9}$ (bottom panels) are reported as a function of the ridge spacing $S$ and ridge width $W$ for $Re_{\tau}=1000$ for elliptical ridges (left column), rectangular ridges (central column) and triangular ridges (right column).}
	\label{fig:scaled}
	\end{figure}

The strength of the secondary flows is scaled using the mean ridge height
\begin{equation*}
\overline{H}=\frac{A_r}{S}    
\end{equation*}
where $A_r$ is the ridge cross section area. 
The maps of the scaled kinetic energy $\mathcal{\hat{K}}$ (top panels), the scaled maximum of the streamfunction $\max_{x_2,x_3} \hat{\psi}^{(1)}$ (central panels) and the scaled dispersive stress $\hat{R}_{12}^{0.9}$ (bottom panels) are here reported. 
The kinetic energy is divided by the square of the mean ridge height $\overline{H}$ as well as the quantity $R_{12}^{0.9}$. Otherwise, the velocity profiles and the streamfunction are scaled with $\overline{H}$. 
Maps of the scaled quantities are compared across the three geometries in figure \ref{fig:scaled}. Note that, when scaling the strength of the secondary flows with the mean ridge height $\overline{H}$, the effective height of the ridges is constant and it is unitary for all cases studied. These maps show strong similarities for the region characterized by high ratio $S/W$ \textit{and} small $W$. We define these two areas of the parameter space as the ``isolated ridge'' regime, where ridges are small in width compared to the distance between one another, and the ``narrow ridge'' regime, where the ridge is narrow compared to the channel half height.

\begin{figure}
    \center
	\includegraphics[width=1\textwidth]{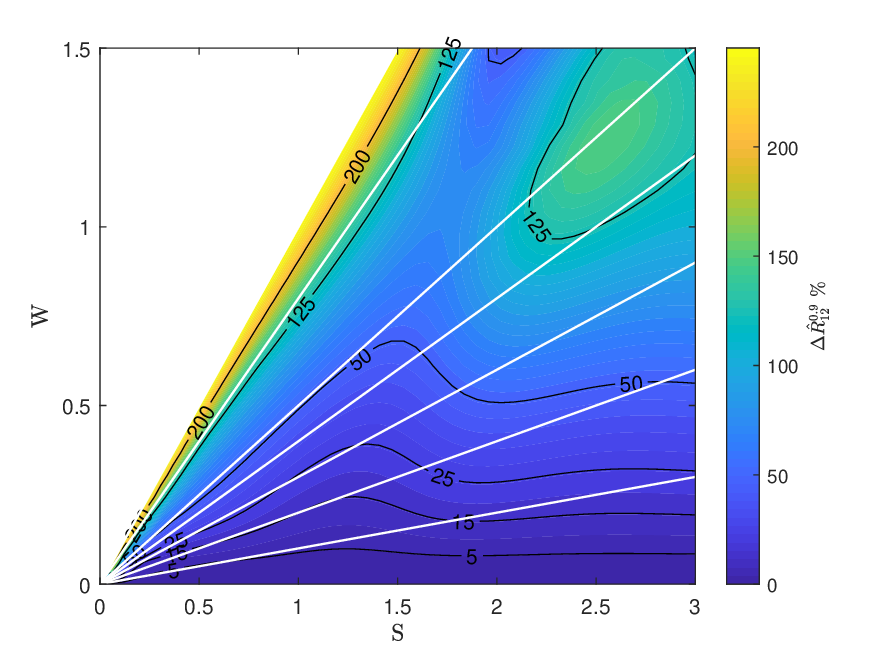}
	\caption{Map of the quantity  is $\Delta \hat{R}_{12}^{0.9} \%$. The black contours are plotted for $\Delta \hat{R}_{12}^{0.9} \%=5,15,25,50,125$ and $200\%$. The white straight lines are obtained for a constant duty cycle $DC=0.1, 0.2, 0.3, 0.4, 0.5$ and 0.8, moving from the bottom to the upper line. }
	\label{fig:error}
	\end{figure}

To better visualise this behaviour, we introduce the quantity
\begin{equation}
\Delta \hat{R}_{12}^{0.9} \%= \frac{\max \hat{R}_{12}^{0.9}- \min \hat{R}_{12}^{0.9}}{\text{mean}\,\hat{R}_{12}^{0.9}} \cdot 100,
\end{equation}
where the function ``$\max$" is the maximum value obtained across the three geometries for a fixed configuration $(S,W)$. Similarly, we define the functions ``$\min$" and ``$\text{mean}$" as the minimum and the mean value, respectively.  The quantity $\Delta \hat{R}_{12}^{0.9} \%$ can be interpreted as the difference in secondary flows strength across the three geometries for the same width and spacing.  This quantity is plotted in figure \ref{fig:error}. We can observe that the relative difference in scaled strength is small if the ridges are narrow and isolated. If this condition is not met, for configurations where the ridges are wide or tightly packed, the difference in the scaled dispersive stresses across the three geometries increases. These differences in the strength of secondary flows observed for elliptical, rectangular and triangular ridges can be easily explained as a consequence of the differences in the flow structures predicted for wide ridges. 
As observed for the flow topology in figure \ref{fig:topology}, tertiary flows emerge for the given $S$ and $W$ only for triangular ridges when the available space between the secondary flows is large enough. 
It is thus interesting a detailed investigation of the reason why these discrepancies between the ridge geometries occur (see section \ref{sec:trapezoidal}).

\section{Scaling the velocity profiles}
\begin{figure*}
	\centering
	\includegraphics[width=\textwidth]{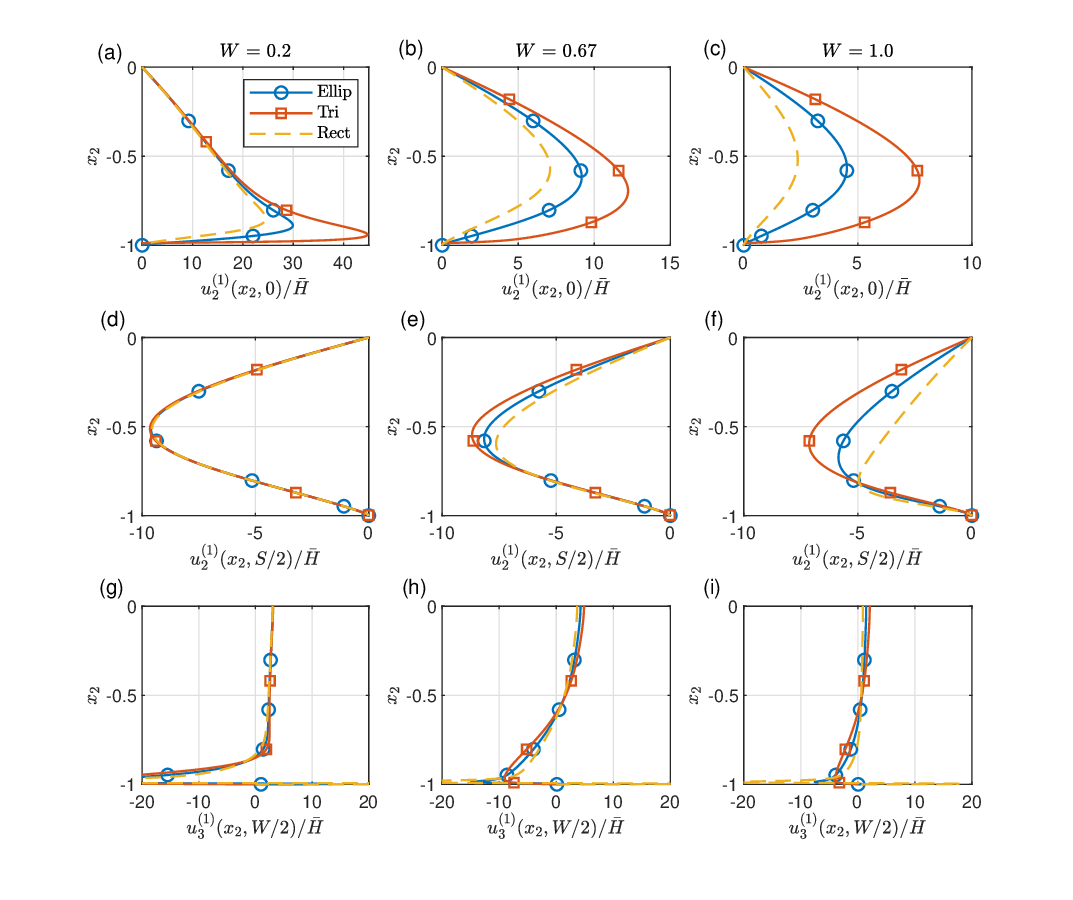}
		\caption{Scaled profiles of the wall-normal velocity component at the centre of the ridge (top panels) and at the centre of the gap (central panels). The spanwise velocity component is extrapolated at the ridge edge (bottom panels).  The spacing is $S=1.25$ and the ridge width $W=0.2$ (left column), $0.67$ (central column) and $1.0$ (right column). The profiles are obtained for the elliptical ``Ellip'', triangular ``Tri'' and rectangular ``Rect'' ridges for $Re_{\tau}=1000$. }
	\label{fig:scaling}
	\end{figure*}

For a better characterization of the scaling of the secondary flows, the velocity components profiles divided by $\overline{H}$ for elliptical, triangular and rectangular ridges are reported in figure \ref{fig:scaling}. Three cases are here considered for a constant spacing $S=1.25$ and varying width in order to display the scaling behaviour for narrow \textit{and} isolated ridges for $W=0.2$ (left column), and the scaling breakdown for $W=0.67$ (central column), corresponding to the maximum amplification configuration, and for  $W=1.0$  (right column), corresponding to wide ridges.
The wall-normal velocity profiles for the three geometries at the ridge centre and at the centre of the trough are reported in the top and central panels, respectively. The spanwise velocity component at the ridge edge is also provided in the bottom panels.
The wall-normal velocity $u_2^{(1)}/\overline{H}$  at the ridge centre (top panels) collapses in the far-wall region only for $W=0.2$ whereas for increasing ridge width the scaled profiles differ and no collapsing is observed. Close to the wall (panel a), the velocity profiles show a peak value that depend on the ridge geometry. This is due to the proximity to the ridge that locally modify the flow field. For increasing width (panel b and c), the velocity profiles differ. The ridge geometry affects the peak value (higher for triangular ridges). For increasing $W$, the strength of the vortices decreases, too.
Similarly, the wall-normal velocity at the centre of the trough is plotted in panels (d,e,f) of figure \ref{fig:scaling}. These velocity profiles collapse for the entire channel height only for the isolated ridge configuration ($W=0.2$). To explain the collapse of the profiles at the centre of the trough for narrow \textit{and} isolated ridges, one can observe that the secondary flows at the centre of the trough are not affected by the ridge geometry, since the distance from the nearest ridge is large compared with $W$. The peak value slightly changes for $W=0.67$. 
For $W=1.0$ where the gap is reduced, the velocity profiles are strongly dependent on the ridge geometry and its effect is not negligible when moving towards the centre of the channel. However, a collapse of the curves is still observed in the near wall region where the influence of the ridges is weak. 
For all three geometries and for $W=0.2$, the spanwise velocity profiles in the bottom panels collapse along the entire wall-normal direction. A negative peak is predicted at the wall over the ridge edge where the Reynolds stresses are stronger. Except for the case $W=0.2$, the peak value depends on the ridge geometry. Moving towards the channel centre, the velocity decreases in magnitude. For increasing width, the scaling breaks down and some differences can be observed across the three geometries.
 
\section{Tertiary flows over trapezoidal ridges} \label{sec:trapezoidal}
In the previous sections, we investigated simple geometries and the flow structure and strength  as a function of the geometrical parameters $S$ and $W$. We observed that secondary flows develop on the ridge edge where there is a strong discontinuity in the ridge shape. This is particularly true for both elliptical and rectangular ridges while for triangular ridges, secondary flows always develop at the ridge crest. Tertiary flows were only observed for the triangular ridges at the given configuration $DC=0.5$ and $W=1$ as shown in figure \ref{fig:topology}. This configuration is characterised by wide and sparse ridges and relies outside the scaling region. Thus, we can conclude that the scaling behaviour does not hold when tertiary flows appear. Trapezoidal ridges are here studied because they combine the main properties of both rectangular and triangular ridges. This allows to better understand how the combination of the shape properties affects the generation of the secondary and tertiary flows. It is worth noting that  for the ridge shapes showing a strong discontinuity, the secondary flows are locked at a ridge edge. 

\begin{figure}
    \centering
    \includegraphics[width=\textwidth]{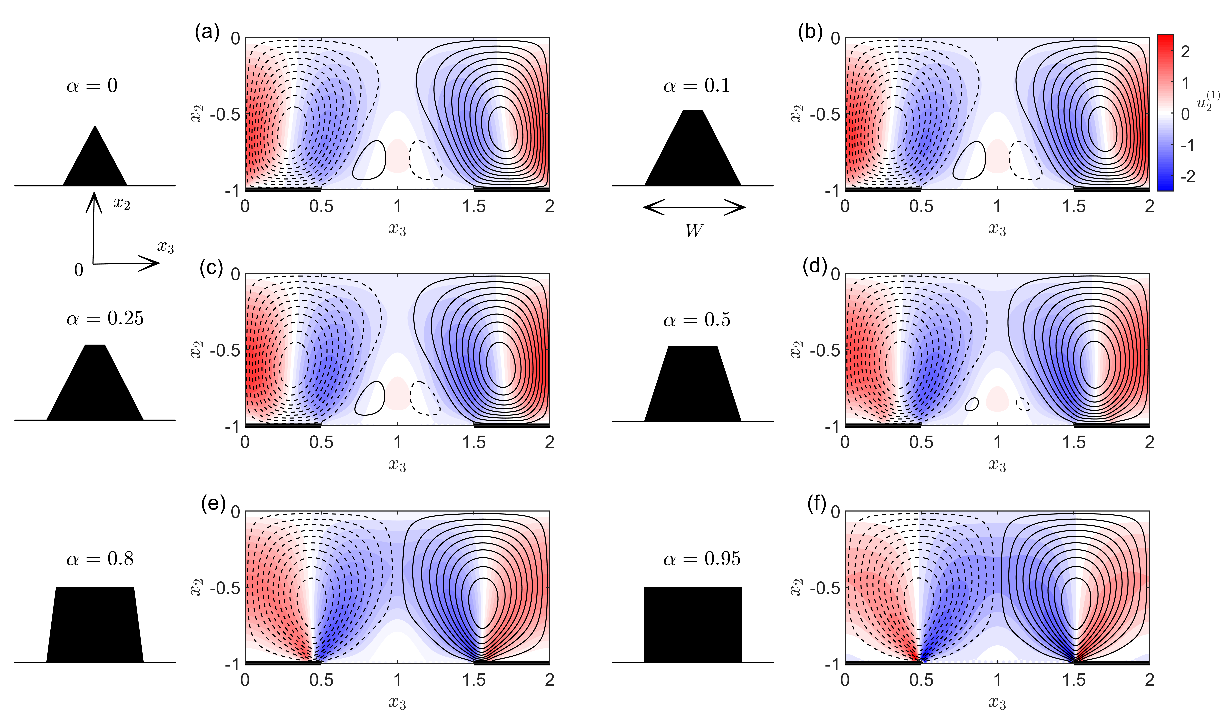}
    \caption{Secondary flows generated by trapezoidal ridges at $Re_\tau=1000$ and for $(S,W)=(2,1)$. The shape parameter $\alpha$ varies from 0, corresponding to a triangular ridge in panel (a), to 0.95, corresponding to a rectangular ridge in panel (f). The value of $\alpha$ is reported. For the sake of clarity, a sketch of the ridge shape is also provided for each configuration studied. }
    \label{fig:trap_topology}
\end{figure}

For this purpose, in addition to the spacing $S$ and the ridge width $W$, we introduce a third geometrical parameter $\alpha$ as the ratio between the minor and major bases of the ridges. A sketch of the ridges considered is shown in figure \ref{fig:trap_topology} on the left-hand side of the corresponding flow topology.
Note that triangular and rectangular ridges are the two limit cases corresponding to $\alpha=0$ and $\alpha=1$, respectively. 
The flow organisation as a function of $\alpha$ is reported in figure \ref{fig:trap_topology}  for $S=1.25$ and $W=0.65$ at $Re_\tau=1000$ (see the sketches on the left). Starting from triangular ridges, tertiary flows develop at the centre of the troughs while the secondary flows 
develop over the flank of the ridge. When $\alpha$ increases, the secondary currents slightly moves towards the ridge edges and the tertiary flows decrease in size and strength, until they disappear at $\alpha=0.95$. 
In particular, tertiary flows emerge when the the space between the secondary vortices is large enough. This also explains why for a fixed configuration, characterised by large spacing and width, different shapes show different flow topologies. Therefore, we can estimate the position, orientation and the size of the secondary vortices for every ridge shape. 

\begin{figure}
    \centering
    \includegraphics[width=\textwidth]{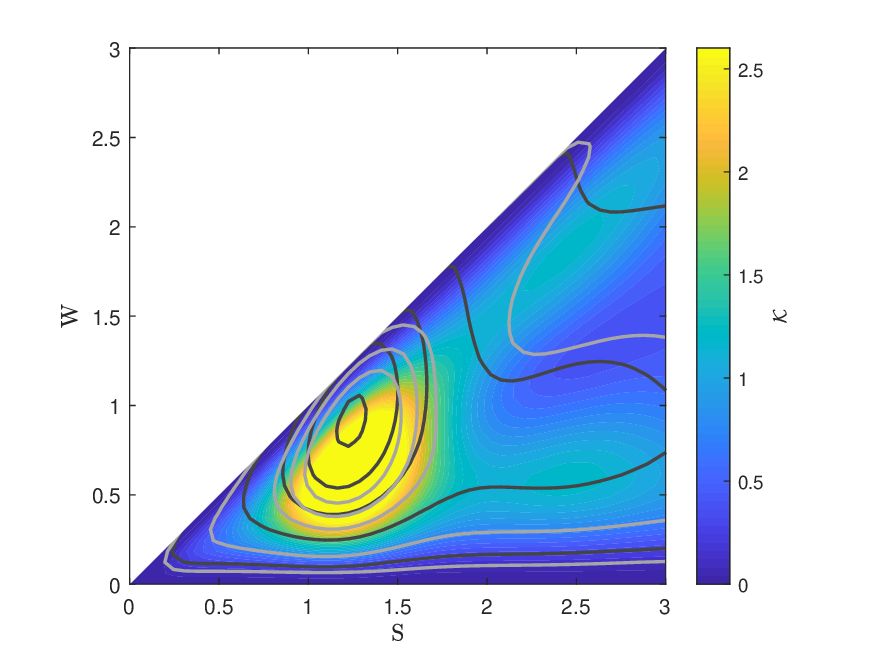}
    \caption{Maps of the kinetic energy density $\mathcal{K}$ of the secondary flows over trapezoidal ridges as a function of the spacing $S$ and width $W$. The colourmap corresponds to the strength of the secondary flows generated by the rectangular ridges. The dark gray lines corresponds to the contour of the triangular ridges while the light gray lines are obtained for the trapezoidal ridges with $\alpha=0.5$. The contour levels are $\mathcal{K}=[0.5,1, 1.5, 2, 2.5]$. The Reynolds number is $Re_\tau=1000$.}
    \label{fig:trap_str}
\end{figure}

The strength of secondary flows is then obtained for the trapezoidal ridges at varying $\alpha$. The kinetic energy density $\mathcal{K}$ is reported in figure \ref{fig:trap_str} where the contours of $\mathcal{K}$ for $\alpha=0.25,0.5$ and 0.8 are superimposed to the colourmap of $\mathcal{K}$ for the rectangular ridges (see the caption for the figure for more details). The peak amplification slightly changes with $\alpha$. For $\alpha=0.25$, the peak amplification occurs for a slightly higher $W$ than the rectangular ridges (black lines).  Also the second peak amplification, observed for a constant width, slightly changes. It is worth noting that for trapezoidal ridges we predict a secondary peak for large width that is not provided for triangular ridges.

\section{Conclusions} \label{sec:conclusions}
The present approach provides predictions of secondary flows produced by different ridge geometries using a linearized RANS based model.
The model allows to quickly predict the flow organization for a large variety of ridge configurations, at a reduced computational cost. Here, secondary flows over elliptical, triangular and rectangular ridges are investigated.
The core assumption is that the flow field over a heterogeneous surface can be decomposed into a base flow and a flow perturbation. In the limit of small ridges, linearized equations are obtained for the perturbation streamwise velocity and the streamfunction perturbation and they are solved by imposing inhomogeneous boundary conditions to model the effect of the ridges.
Since the equations are linear, the superposition principle is applied to reconstuct the flow field over complex geometries.   Only the results for $Re_\tau=1000$ are reported. 
 
Strong similarities in the configuration of the secondary flows are observed for narrow ridges with small width compared to the spacing. For increasing width, the vortices grow in strength and size until they occupy the channel half-height. The flow organization differs between the three geometries for wide ridges where the secondary structures depend on the ridge geometry.
 
 The similarity in the flow topology  for narrow ridges suggest that the velocity induced has been scaled using  an appropriate parameter that depends on the ridge geometry. Since the ridges are localized at the wall, the effect of the wall perturbation is assumed to be proportional to the cross-sectional area of the ridges, not their geometry. For this reason, the mean ridge height $\overline{H}=A_r/S$ is here proposed as the scaling parameter.
 
 In particular the maps of the volume averaged dispersive stresses $R_{12}^{0.9}$ are 
 are scaled with $\overline{H}^2$ and they collapse for the ridge configurations with small width $W$, defining the ``narrow ridges" regime, \textit{and} high ratio $S/W$, defining the ``isolated ridges" regime. 
 For these configurations the effect of the ridge geometry is negligible or confined to the very near-wall region and the scaled velocity profiles collapse. 
 In particular, the collapse of the profiles at the centre of the trough can be explained as the consequence of the low influence of the nearest ridge. For increasing width, the scaling behaviour breaks down because the flow topology strongly depends on the ridge geometry. 

Finally, the generation mechanism of the tertiary flows is studied for trapezoidal ridges.
The present paper confirms that tertiary flows emerge when the space between the secondary flows is large enough. These results extend the conclusions about the generation of tertiary flows for rectangular ridges by \citet{zampino2022} to a more general shape.

\bibliographystyle{unsrtnat}
\bibliography{references}  






\end{document}